# Scattering-dependent transport of SrRuO$_3$ films: From Weyl fermion transport to hump-like Hall effect anomaly

Shingo Kaneta-Takada,[1,2,*] Yuki K. Wakabayashi,[1,*,†] Yoshiharu Krockenberger,[1] Hiroshi Irie,[1] Shinobu Ohya,[2,3] Masaaki Tanaka,[2,3] Yoshitaka Taniyasu,[1] and Hideki Yamamoto[1]

[1]*NTT Basic Research Laboratories, NTT Corporation, Atsugi, Kanagawa 243-0198, Japan*
[2]*Department of Electrical Engineering and Information Systems, The University of Tokyo, Bunkyo, Tokyo 113-8656, Japan*
[3]*Center for Spintronics Research Network (CSRN), The University of Tokyo, Bunkyo, Tokyo 113-8656, Japan*

[*]These authors contributed equally to this work.
[†]Corresponding author: yuuki.wakabayashi.we@hco.ntt.co.jp

**Abstract**

Recent observation of quantum transport phenomena of Weyl fermions has brought much attention to 4$d$ ferromagnetic perovskite SrRuO$_3$ as a magnetic Weyl semimetal. Besides, the hump-like Hall effect anomaly, which might have a topological origin, has also been reported for this material. Here, we show that the emergence of such phenomena is governed by the degree of scattering determined by the defect density (Ru-deficiency- and/or interface-driven-defect scattering) and measurement temperature (phonon scattering), where the former is controlled by varying the growth conditions of the SrRuO$_3$ films in molecular beam epitaxy as well as the film thickness. The resulting electronic transport properties can be classified into three categories: clean, intermediate, and dirty regimes. The transport of Weyl fermions emerges in the clean regime, whereas that of topologically trivial conduction electrons in the ferromagnetic metal state prevail in the intermediate and dirty regimes. In the clean and intermediate regimes, anomalous Hall resistivity obeys a scaling law incorporating the intrinsic Karplus-Luttinger (Berry phase) and extrinsic side-jump mechanisms. The hump-like Hall effect anomaly is observed only in the dirty regime, which is contrary to the scaling law between anomalous Hall resistivity and longitudinal resistivity. Hence, we conclude that this anomaly is not inherent to the material and does not have a topological origin. We also provide defect- and temperature-dependent transport phase diagrams of stoichiometric SrRuO$_3$ and Ru-deficient SrRu$_{0.7}$O$_3$ where the appearance of Weyl fermions and hump-like Hall effect anomaly are mapped. These diagrams may serve as a guideline for designing SrRu$_{1-x}$O$_3$-based spintronic and topological electronic devices.

# I. INTRODUCTION

Topological materials have attracted much attention for future applications in spintronics and topological electronics [1]. Novel functional devices such as racetrack memory [2,3] utilizing skyrmions [4] and topoelectrical circuits [5,6] utilizing Weyl fermions have been proposed [7,8]. Weyl fermions emerge at the spin-split band crossing points in momentum space with spin-orbit interaction [9−13]. They show novel transport phenomena related to pairs of Weyl nodes, such as chiral-anomaly-induced negative magnetoresistance (MR), unsaturated linear positive MR, the $\pi$ Berry phase along cyclotron orbits, light cyclotron masses, and high quantum mobility [14]. Meanwhile, skyrmions are topologically nontrivial spin textures derived from a correlation between a strong Dzyaloshinskii-Moriya interaction and a ferromagnetic exchange interaction. A skyrmion creates the Berry curvature and provides additional transverse electron scattering, leading to the hump-like Hall effect anomaly, the so-called topological Hall effect [15].

In recent years, itinerant $4d$ ferromagnetic perovskite $SrRuO_3$ has attracted intense interest due to the emergence of Weyl fermions in oxides [16,17] and the hump-like Hall effect anomaly that possibly arises from the topologically nontrivial spin texture [18−20]. The existence of Weyl fermions is well established, and high-quality $SrRuO_3$ films show quantum transport of bulk three-dimensional Weyl fermions and two-dimensional Weyl fermions from surface Fermi arcs [21,22]; in our previous studies [17,21,23], we observed the five signatures of Weyl fermions described above in quantum transport in $SrRuO_3$. The presence of the Weyl fermions in $SrRuO_3$ has also been confirmed by first-principles calculations and angle-resolved photoemission spectroscopy (ARPES) observations of linear band dispersions [24]. In contrast, the origins of the hump-like Hall effect anomaly are still controversial [25−27]. Various mechanisms have been proposed and discussed, including skyrmion-driven topological Hall effect [18−20] and multi-channel anomalous Hall effect (AHE) [28−31]. The latter originates from the superposition of AHEs with different signs of hysteresis loops when non-identical multiple magnetic domains exist. Here, the AHE is an additional contribution to the Hall resistivity due to the localized magnetic moment characteristic of magnetic materials.

It is of crucial importance here to draw a distinction between effects that are specific to $SrRuO_3$ and those of accompanying nature. The skyrmion-driven topological Hall effect is a topological and intrinsic effect while the multi-channel AHE is a non-topological and extrinsic one. When the hump-like Hall effect was first observed in $SrRuO_3$, it was attributed to the topological Hall effect arising from the skyrmions generated at the $SrRuO_3/SrIrO_3$ interface [18,19]. On the other hand, the hump-like Hall effect anomaly has been observed when $SrRuO_3$ is composed of magnetically inhomogeneous domains [28−31]. Although Ru deficiencies are considered essential for the inhomogeneity of magnetic domains [28−31], the Ru-deficiency-dependent manifestation of the hump-like Hall effect anomaly has not been systematically investigated, because of the difficulty of controlling the amount of Ru deficiency in $SrRuO_3$. Furthermore, for single layer $SrRuO_3$, this phenomenon is only observed in very thin films (< ~6 nm) presumably with interface-driven defects [20,28−36]. Therefore, systematic experiments on the Weyl fermion transport and hump-like Hall effect anomaly in stoichiometric and Ru-deficient films with various thicknesses are required to distinguish topologically trivial or nontrivial transport phenomena.

In this study, we investigated the magnetotransport properties of epitaxial SrRuO$_3$ and SrRu$_{0.7}$O$_3$ films grown by machine-learning-assisted molecular beam epitaxy (ML-MBE) [37,38], in which the supply ratios of Sr and Ru during deposition can be controlled at designated values at will. We prepared stoichiometric (SrRuO$_3$) and Ru-deficient (SrRu$_{0.7}$O$_3$) films with various thicknesses $t$ (= 2–60 nm). We found that a variety of magnetotransport properties that those films show at different temperatures can be classified into three categories, namely those observed in clean, intermediate, and dirty regimes. For example, Weyl fermion transport is observed only in the clean regime, *i.e.*, in the thick ($\geq$ 10 nm) stoichiometric films at low temperatures; on the other hand, the hump-like Hall effect anomaly is observed only in the dirty regime, *i.e.*, in the thin ($\leq$ 5 nm) Ru-deficient films. We also examined a scaling law between anomalous Hall resistivity and longitudinal resistivity, which is derived from a model incorporating the intrinsic Karplus-Luttinger (Berry phase) and extrinsic side-jump mechanisms. While this scaling law can be applied broadly to most of the SrRuO$_3$ and SrRu$_{0.7}$O$_3$ films in the clean and intermediate regimes, regardless of stoichiometry, film thickness, and temperature, we found it not the case for those exhibiting the hump-like Hall effect anomaly (in the dirty regime). We also provide defect- and temperature-dependent transport diagrams for SrRuO$_3$ and SrRu$_{0.7}$O$_3$ for accessing various electronic states realized in these materials, including the topological one with Weyl fermions and the non-topological one that exhibits the hump-like Hall effect anomaly. The diagrams will serve as a guideline for designing SrRu$_{1-x}$O$_3$-based spintronic and topological electronic devices.

## II. EXPERIMENTS

We grew epitaxial stoichiometric SrRuO$_3$ and Ru-deficient SrRu$_{0.7}$O$_3$ films with various $t$ (= 2–60 nm) on (001) SrTiO$_3$ substrates in a custom-designed ML-MBE setup equipped with multiple e-beam evaporators for Sr and Ru [39−41]. For the SrRuO$_3$ films, the growth parameters were optimized by Bayesian optimization [37,38,42], a machine learning technique for parameter optimization. We precisely controlled the elemental fluxes, even for elements with high melting points, e.g., Ru (2250 °C), by monitoring the flux rates with electron-impact-emission-spectroscopy, and the flux rates are fed back to the power supplies for the e-beam evaporators [41,43]. The Ru flux rates were set at 0.190 and 0.365 Å/s for the growth of the Ru-deficient and stoichiometric films, respectively, while the Sr flux was kept at 0.98 Å/s. The supplied Ru rates of 0.190 Å/s for Ru-deficient and 0.365 Å/s for stoichiometric films correspond to the Ru/Sr flux ratios of 0.80 (Ru-poor) and 1.54 (Ru-rich), respectively [23,44]. Excessive Ru is known to re-evaporate from the growth surface by forming volatile species such as RuO$_4$ and RuO$_3$ under oxidizing atmospheres, and this thermodynamic behavior has been exploited for stoichiometric films [45]. The oxidation during growth was carried out with a mixture of ozone (O$_3$) and O$_2$ gas (~15% O$_3$ + 85% O$_2$), which was introduced at a flow rate of ~2 sccm through an alumina nozzle pointed at the substrate. All SrRuO$_3$ and SrRu$_{0.7}$O$_3$ films were grown at 772°C. The chemical compositions of the SrRuO$_3$ and SrRu$_{0.7}$O$_3$ films with thickness $t$ = 60 nm were determined using energy dispersive X-ray spectroscopy (EDS) [44]. The Ru deficiency in the SrRu$_{0.7}$O$_3$ film was also confirmed by the chemical shift in the Ru 3$p_{3/2}$ X-ray photoelectron spectroscopy (XPS) peak [44]. Further information about the MBE setup, preparation of the substrates, reflection high-energy electron diffraction patterns, cross-sectional scanning transmission electron microscopy images, surface morphology, X-ray diffraction patterns, EDS, and XPS of stoichiometric SrRuO$_3$ and Ru-deficient SrRu$_{0.7}$O$_3$ films are described elsewhere [17,21,23,37−49].

For the measurement of transport properties, we fabricated 200 × 350 μm$^2$ Hall bar structures by photolithography and Ar ion milling. Before making the Hall bar structure, we deposited Ag electrodes onto SrRuO$_3$ or SrRu$_{0.7}$O$_3$ surfaces. The longitudinal resistivity $\rho_{xx}(T)$ data in Refs. [23] and [44] were obtained from the same samples used in this study, but without making the Hall bar. The magnetotransport up to 14 T in the temperature range of 2 to 300 K was measured in a DynaCool physical property measurement system (PPMS). The magnetic field $B$ was applied in the [001] direction of the SrTiO$_3$ substrate perpendicular to the film plane. The current flow in the specimens is nearly along the in-plane [100] direction of the pseudo-cubic SrRuO$_3$ and SrRu$_{0.7}$O$_3$.

### III. RESULTS AND DISCUSSION

We performed magnetotransport measurements on the Hall bar devices of the stoichiometric SrRuO$_3$ and Ru-deficient SrRu$_{0.7}$O$_3$ films with $t$ = 2–60 nm. The temperature ($T$) dependence of the longitudinal resistivity $\rho_{xx}(T, 0\ \text{T})$, residual resistivity ratio (RRR), and the Curie temperature $T_C$ of all the films are summarized in Fig. S1 of Supplemental Material [46]. In short, we achieved a high RRR > 60 when the film thickness of the stoichiometric film was large ($t$ = 60 nm) [17,21,23,37,38,44–49]. The RRR value is a measure of the purity thus providing information on the degree of impurity (defect) scattering. For RRR as high as > 60, SrRuO$_3$ is regarded as a high-quality material. For SrRuO$_3$ films with $t$ = 5 − 60 nm and SrRu$_{0.7}$O$_3$ films with $t$ = 10 − 60 nm, temperature dependencies of $\rho_{xx}(T, 0\ \text{T})$ were almost completely preserved after the Hall-bar fabrication although the absolute values are ~ 1.3 times higher, confirming limited fabrication damage, in addition to the high reproducibility of our experiments [Fig. S1(a), S1(b), and Refs. 23 and 44].

### A. Magnetoresistance (MR)

Figure 1 shows the $t$ dependence of the MR [($\rho_{xx}(B)−\rho_{xx}(0\ \text{T}))/\rho_{xx}(0\ \text{T})$] of SrRuO$_3$ [Figs. 1(a) and 1(b)] and the SrRu$_{0.7}$O$_3$ [Figs. 1(c) and 1(d)] films when $B$ was applied in the out-of-plane [001] direction of the SrTiO$_3$ substrate at 2 and 60 K. For stoichiometric SrRuO$_3$ films at 2 K, the sign of MR in the high-$B$ region changed from negative to positive with increasing $t$ [Fig. 1(a)]. The positive linear MR for $t \geq 10$ nm at 2 K showed no signature of saturation even up to 14 T, which is commonly seen in Weyl semimetals and is thought to stem from the linear energy dispersion of Weyl nodes [14]. Furthermore, these SrRuO$_3$ films with $t \geq 10$ nm showed quantum oscillations in resistivity (*i.e.*, SdH oscillations), whose frequency (26 T) corresponds to that of the Fermi arc of Weyl fermions with high mobility of 3,000 to 10,000 cm$^2$/Vs [17,21,23]. As described in the introduction, these unsaturated linear positive MR and SdH oscillations, which appear simultaneously when $t \geq 10$ nm, are incontrovertible facts of Weyl fermion transport in SrRuO$_3$. Accordingly, Weyl fermions are playing a vital role also in SrRuO$_3$ films processed into the Hall-bar structure. In case of Ru-deficiency, *i.e.*, SrRu$_{0.7}$O$_3$ films, however, the MR in the high-$B$ region is negative, and the SdH oscillations are not observed for any film even at 2 K irrespective of the thickness [Fig. 1(c)]. In other words, electronic transport by Weyl fermions may not be dominant, which is consistent with our previous report [44].

At 60 K, all of the samples show a negative MR due to the suppression of magnetic scattering at high magnetic fields [17], irrespective of thickness and stoichiometry [Figs. 1(b) and 1(d)]. In all of the films, the anisotropic MR (AMR), which is proportional to the relative angle between the electric current and the magnetization, show a MR hysteresis where the peak position of the MR corresponds to the coercive field $H_C$. $SrRu_{0.7}O_3$ films with $t = 20-60$ nm show a rather peculiar AMR shape, and the MR hysteresis appears above ~1 T [Figs. 1(c) and 1(d)]. This feature has been attributed to the existence of two different magnetic components in the $SrRu_{0.7}O_3$ films [44]. The AMR disappears above $T_C$, reflecting the paramagnetic states [Figs. S2(c) and S2(d)].

### B. The Hall effect

Figures 2(a) and 2(b) [2(c) and 2(d)] show the $B$ dependence of the Hall resistivity $\rho_{xy}(B)$ for the stoichiometric $SrRuO_3$ [Ru-deficient $SrRu_{0.7}O_3$] films at 2 and 60 K, well below the $T_C$ values of each film [Fig. S1(f)]. The $\rho_{xy}(B)$ curves are sectioned into three categories based on their qualitative characteristics: (1) the clean regime where $SrRuO_3$ films with $t \geq 10$ nm (RRR > ~20) show the Weyl fermion transport at low temperatures below ~20 K, (2) the intermediate regime where the Weyl fermion transport is suppressed and the topologically trivial conduction electrons in the ferromagnetic metal govern the transport phenomenon, (3) the dirty regime where the hump-like Hall effect anomaly emerges in the Ru-deficient $SrRu_{0.7}O_3$ films with $t \leq 5$ nm (RRR < 3.5).

#### 1. Clean regime: $SrRuO_3$ with $t = 10-60$ nm measured at 2K

The $\rho_{xy}(B)$ curves for the stoichiometric $SrRuO_3$ films with $t = 10$-60 nm at 2 K are varying monotonically with negative (positive) slopes at low (high) $B$ ranges. This non-linear behavior is reminiscent of semimetals with multiple carrier types [Fig. 2(a)]. In ultra-high-quality $SrRuO_3$ films with $t = 10-60$ nm, Weyl fermion transport dominates at 2 K, and accordingly, the observed $\rho_{xy}(B)$ behavior is assumed to be determined by Weyl fermions in the Weyl semimetal state. This is further supported by the Shubnikov-de Haas (SdH) oscillations superimposed in the $\rho_{xy}(B)$ curves [see inset of Fig. 2(a) as an example], which can be attributed to two-dimensional Weyl fermion transport from surface Fermi arcs [21]. This in turn indicates that the $B$ dependence of Hall resistivity characteristic of semimetals may be utilized as another signature of the Weyl fermion transport in $SrRuO_3$.

Although Weyl fermions in solids follow a linear dispersion relation while there are several Weyl points around the Fermi energy ($E_F$) within a single Brillouin zone of $SrRuO_3$ [17], a discussion using a two-carrier model assuming two (one convex-upward and the other downward) quadratic dispersions would provide some clues to the interpretation of the $\rho_{xy}(B)$ data [50]. According to the two-carrier model, the sign and the slope of $\rho_{xy}(B)$ in the high-$B$ limit reflect the major carrier type and the difference between the major-carrier and minor-carrier densities, respectively. The positive sign in the high-$B$ range is indicative for hole-like Weyl fermions. The negative slope in the low-$B$ range may reflect on the fact that $\mu_e$ is higher than $\mu_h$ because $p$ is higher than $n$ [50]. Here, $p$, $n$, $\mu_h$, and $\mu_e$ are the hole carrier density, electron carrier density, hole mobility, and electron mobility, respectively. While topologically trivial conduction electrons in the metallic state of $SrRuO_3$ may also contribute to the observed Hall resistivity, the striking contrast between the data for the stoichiometric $SrRuO_3$ films with 10−60 nm measured at 2 K and those measured at 60 K [Figs. 2(a) and 2(b)]—*e.g.*, the slope is positive for the former and negative for the latter in the high-$B$ ranges—suggests that the Weyl fermion transport prevails at 2 K. The nonlinear-$B$ behavior of $\rho_{xy}(B)$ together with the oversimplified two-carrier model render an exact estimation of the carrier densities inaccurate, the slope of $\rho_{xy}(B)$ in the high-$B$ range gives a carrier density in the range of $10^{22}$ cm$^{-3}$. This is much larger than that estimated from the SdH oscillations (~$10^{17}$ − $10^{18}$ cm$^{-3}$) [17], where only charge carriers that complete cyclotron orbits during their lifetimes are counted. While the monotonic variation of $\rho_{xx}(T, 0\text{ T})$ with decreasing $T$ for $T \leq 60$ K [Fig. S1(a)] imply no dramatic change of the majority charge carriers from the topologically trivial conduction electrons to the Weyl fermions but rather a crossover, Weyl fermion transport is prominent in measurements under magnetic fields such as those of the Hall effect, SdH oscillations, and MR at low temperatures.

*2. Intermediate regime: SrRuO$_3$ with t = 2−5 nm measured at 2K, SrRuO$_3$ with t = 2 −60 nm measured at 60 K, and SrRu$_{0.7}$O$_3$ with t = 10−60 nm measured at 2 or 60 K*

The transport properties in this category are subject to substantial scattering due to defects and/or phonons. The defect-induced scattering is characterized by low residual resistivity ratio and the defects are interface- and/or Ru-deficiency-driven. The scattering events take place too frequently and Weyl fermion transport and SdH oscillations are marginalized (Fig. 2) [44]. Thus, the observed Hall effect can be attributed to the topologically trivial conduction electrons in the metallic SrRuO$_3$ or SrRu$_{0.7}$O$_3$. In the intermediate regime, magnetization hysteresis curves characteristic of AHE are clearly observed, and it dominates Hall resistivity at near-zero magnetic fields ($\rho_{xy,0\text{ T}}$). This is quite in contrast to the clean regime, where the AHE components are negligibly small due to the small $\rho_{xx}$ values; $\rho_{xy,0\text{ T}}$ is proportional to $\rho_{xx}^2$ at low enough temperatures. Above $T_C$ (at 180 K), $\rho_{xy,0\text{ T}}$ of all the films becomes zero, reflecting the paramagnetic states [Fig. S2(a)]. The temperature dependence of the AHE will be discussed in detail later [§III. C] with regard to the scaling law between $\rho_{xy}$ and $\rho_{xx}$.

The ordinary Hall effect [linear $\rho_{xy}(B)$] at high $B$, in metals depends on the energy band dispersion [*e.g.*, $dE(\mathbf{k})/d\mathbf{k}$] near $E_F$ and also on $\mathbf{k}$-dependent (anisotropic) relaxation times [$\tau(\mathbf{k})$] [51]. Hence, the interpretation of the data is not straightforward. More so, SrRuO$_3$ has spin-polarized electron bands for $T < T_C$, and its Fermi surface consist of at least six sheets, including open orbits [52]. Accordingly, the Hall coefficient is not related to the carrier density in any simple manner [51]. Nonetheless, the identical $\rho_{xy}(B)$ slopes in the data except for very thin films (2−5 nm) indicate that the electronic structure, $\tau(\mathbf{k})$, and the carrier density are comparable among the films assigned to the intermediate regime. Supposing that $\tau(\mathbf{k})$ on one of the Fermi sheets (closed and isotropic) is significantly longer than the others and governs the Hall effect, the slope of $\rho_{xy}(B)$ would provide an effective carrier (electron) density of about $2 \times 10^{22}$ cm$^{-3}$ [50]; the corresponding charge carrier mobility at 60 K would be 7−9 cm$^2$/Vs and 3−4 cm$^2$/Vs for SrRuO$_3$ and SrRu$_{0.7}$O$_3$ with $t$ = 10−60 nm, respectively [Fig. S3]. Unexpectedly, the slopes of $\rho_{xy}(B)$ in the high-$B$ range for stoichiometric and Ru-deficient films ($t$ = 10−60 nm) at 60 K are identical nevertheless the band filling factor should be different between SrRuO$_3$ and SrRu$_{0.7}$O$_3$ within a rigid band model. This might be explained by assuming a common Fermi surface that mostly governs the Hall effect whose geometry is unchanged by Ru vacancies, while only filling factors of bands with substantially shorter $\tau(\mathbf{k})$ at $E_F$ vary with the Ru composition.

### 3. Dirty regime: SrRu$_{0.7}$O$_3$ with t = 2−5 nm measured at 2K or 60 K

The AHEs with hump structures are prominent in the Ru-deficient films with $t = 2$ and 5 nm at 2 K [Fig. 2(c)]. At 60 K, this hump-like Hall effect anomaly for $t = 2$ nm disappears, while persisting for $t = 5$ nm [Fig. 2(d)]. Previous studies have suggested that these hump-like Hall effect anomalies in SrRuO$_3$ are associated to the topological Hall effect [18−20] and the multi-channel AHEs from the superposition of two different AHE components [28−31]. In our films, the humps are observed only for Ru-deficient SrRu$_{0.7}$O$_3$ thin films, indicating that Ru-deficiencies play an important role in the appearance of the humps; *i.e.*, it does not have an intrinsic nor topological origin. Instead, it is known that this material shows different signs of AHE depending on $T$ [53]. The net AHE can be accompanied by a hump structure if the temperature at which its sign changes differs among some multiple magnetic domains.

### C. Further analysis of AHE: quest for the origin

Various theoretical models have been proposed for the origin of AHE in magnetic materials, including the Karplus-Luttinger (Berry phase) mechanism [54,55] as an intrinsic origin and the side-jump [56,57] and skew-scattering mechanisms [58,59] as extrinsic origins. To examine the origin of the AHE in our SrRuO$_3$ and SrRu$_{0.7}$O$_3$ films, we measured the temperature dependence of the AHE. Figures 3(a) and 3(b) show the $T$ dependence of the anomalous Hall resistivity $\rho_{xy}$ at 0 T for stoichiometric SrRuO$_3$ and Ru-deficient SrRu$_{0.7}$O$_3$ films. Here, $\rho_{xy,0\,T}$ is the $\rho_{xy}$ value at 0 T after sweeping from high $B$ (9 or 14 T) to 0 T (Fig. S4). The position of kinks of $\rho_{xx}(T)$ curves (Fig. S1) are used to determine $T_C$. Irrespective of stoichiometry and thickness, $\rho_{xy,0\,T}$ becomes zero above $T_C$ [Fig. S2(a) and S2(b)], reflecting the paramagnetic states [60]. Except for the SrRu$_{0.7}$O$_3$ films with $t = 2$ and 5 nm, the $\rho_{xy,0\,T}$ curves show positive peaks around $T_C$, and then the values change from positive to negative with decreasing $T$. For even lower temperatures, $\rho_{xy,0\,T}$ values increase. This trend has been attributed to a combination of the intrinsic Karplus-Luttinger (Berry phase) and extrinsic side-jump mechanisms [53,55−57]. For the scaling analysis of $\rho_{xy,0\,T}$ to $\rho_{xx}$, we defined $\rho_{xy,0\,T}'$ and $\rho_{xx}'$ as $\rho_{xy,0\,T}$ normalized by its maximum absolute value and $\rho_{xx}$ normalized by its value when $\rho_{xy,0\,T}$ changes its sign, respectively. Figure 3(c) shows the $\rho_{xy,0\,T}'$ vs. $\rho_{xx}'$ curves for the SrRuO$_3$ films with $t = 2-60$ nm and the SrRu$_{0.7}$O$_3$ films with $t = 10-60$ nm. The $\rho_{xy,0\,T}'$ vs. $\rho_{xx}'$ curves for the Ru-deficient SrRu$_{0.7}$O$_3$ films with $t = 2$ and 5 nm are also shown in Fig. S5. Regardless of stoichiometry, thickness, and temperature at which $|\rho_{xy,0\,T}|$ becomes maximum, all $\rho_{xy,0\,T}'-\rho_{xx}'$ curves well coincide, indicating that a universal scaling law exists, independently of scattering frequency and sources; *i.e.*, Ru-deficiency- or interface-driven defects, or phonons. Yet, Ru-deficient SrRu$_{0.7}$O$_3$ films with $t = 2$ and 5 nm that show the hump-like Hall effect anomaly [Fig. 2(c) and 2(d)] deviate from this trend (Fig. S5). This indicates that the hump-like Hall effect anomaly cannot be explained by the Karplus-Luttinger or side-jump mechanisms. More importantly, this anomaly is not inherent to the material system of SrRu$_{1-x}$O$_3$. We fit $\rho_{xy,0\,T}'-\rho_{xx}'$ curves using the scaling law expressed as follows [53]:

$$\rho'_{xy,0\,T} = \frac{a_1}{\Delta^2 + a_2(\rho'_{xx})^2}(\rho'_{xx})^2 + a_3(\rho'_{xx})^2, \quad (1)$$

where $a_1-a_3$ and $\Delta$ are fitting parameters, which are related to the band structure of SrRuO$_3$ [53]. The first term describes the contribution from the Karplus-Luttinger mechanism, and the second term that from the side-jump scattering. As shown in Fig.

3(c), all of our experimental data are well fitted by a single set of parameters: $a_1 = -2.3 \times 10^{16}$, $a_2 = 1.3 \times 10^{15}$, $a_3 = 7.4$, and $\Delta = 4.2 \times 10^7$. This universal scaling of the AHE in SrRuO$_3$ films with $t = 2-60$ nm and the SrRu$_{0.7}$O$_3$ films with $t = 10-60$ nm (in the clean and intermediate regimes) supports a common origin from combined Karplus-Luttinger (Berry phase) and side-jump mechanisms. From Fig. 3(c), in the region of $\rho_{xx}' < 1.0$ ($\rho_{xy,0\,T}' < 0$), the contribution from the Karplus-Luttinger mechanism is dominant. On the other hand, in the region of $\rho_{xx}' > 1.0$ ($\rho_{xy,0\,T}' > 0$), the contribution from the side-jump scattering is dominant.

### D. RRR and temperature dependent transport diagram

Finally, we summarize the magnetotransoprt results on RRR-temperature diagrams (Fig. 4) and depict the regions where the Weyl fermion transport or the hump-like Hall effect anomaly emerges. Again, the RRR is known to be an important measure of the crystal purity, *i.e.*, defect density, which was controlled by Ru deficiencies and/or interface defects. The transport phase diagram for stoichiometric SrRuO$_3$ [Fig. 4(a)] is consistent with what we have shown before [17] except for the very low RRR ($\leq 10$) regime. First, the Weyl fermion transport appears only if RRR is high and $T$ is low [Fig. 4(a)]. This observation is physically reasonable as Weyl fermion transport is vulnerable to scattering. Second, the hump-like Hall effect anomaly emerges only if RRR is low [Fig. 4(b)]. More specifically, the hump-like Hall effect anomaly is observed in Ru-deficient SrRu$_{0.7}$O$_3$ films with $t \leq 5$ nm (RRR < ~3.5) but not in the 2-nm thick stoichiometric SrRuO$_3$ film (RRR ~ 2.4). This indicates that the hump-like Hall effect anomaly is caused by the coexistence of Ru deficiencies and interface-driven defects. In fact, this anomaly has been reported only for very thin films (< 6 nm) [20,28−36]. A considerable amount of defects is a prerequisite for the emergence of the hump-like Hall effect anomaly, and it does not have an intrinsic or topological origin.

### IV. SUMMARY

In summary, we have investigated the magnetotransport properties of epitaxial stoichiometric SrRuO$_3$ and Ru-deficient SrRu$_{0.7}$O$_3$ films with various thicknesses ($t = 2-60$ nm). We found that the emergence of Weyl fermion and hump like Hall effect anomaly is governed by the degree of scattering. With respect to scattering, we categorized magnetotransport properties into clean, intermediate, and dirty regimes.

(1) In the clean regime, SrRuO$_3$ films with $t \geq 10$ nm (RRR > ~20) show the Weyl fermion transport (linear positive MR, SdH oscillations, and Hall effect characteristic of semimetals) at low temperatures below ~20 K.
(2) In the intermediate regime, the Weyl fermion transport disappears and the topologically trivial conduction electrons in the ferromagnetic metal govern the transport phenomenon; the ordinary Hall effect is observed accompanied by the anomalous Hall effect (AHE). The ordinary Hall effect suggests that the electronic structure and Fermi surface geometry are common among the SrRuO$_3$ and SrRu$_{0.7}$O$_3$ films classified in this category. We analyzed the correlation between the anomalous Hall resistivity [$\rho_{xy,0\,T}(T)$] and the longitudinal resistivity [$\rho_{xx}(T)$] using a combination of the intrinsic Karplus-Luttinger (Berry phase) and the extrinsic side-jump mechanisms. This results in a universal scaling law for the clean and intermediate regimes, independently of scattering frequency and sources.
(3) In the dirty regime, the hump-like Hall effect anomaly emerges in the Ru-deficient

SrRu$_{0.7}$O$_3$ films with $t \leq 5$ nm (RRR < 3.5). The scaling law between $\rho_{xy,0\,T}$ and $\rho_{xx}$ established for the clean and intermediate regimes does not hold in the dirty regime, indicating that the hump-like Hall effect cannot be explained by the adopted model and is not inherent to SrRu$_{1-x}$O$_3$.

The defect- and temperature-dependent transport diagrams of SrRuO$_3$ and SrRu$_{0.7}$O$_3$ may serve as a guideline for the development of SrRu$_{1-x}$O$_3$-based spintronic and topological electronic devices that exploit various electronic and magnetic states.

## Data Availability

The data that support the findings of this study are available from the corresponding author upon reasonable request.

## Acknowledgements

S.K.T. acknowledges the support from the Japan Society for the Promotion of Science (JSPS) Fellowships for Young Scientists.

## Author Contributions

Y.K.W. conceived the idea, designed the experiments, and directed and supervised the project. Y.K.W. and Y.Kro. grew the samples. S.K.T., Y.K.W., and H.I. fabricated the Hall bar structures. S.K.T. and Y.K.W. carried out the magnetotransport measurements. S.K.T. and Y.K.W analyzed and interpreted the data. S.K.T. and Y.K.W. co-wrote the paper with input from all authors.

## Competing interests

The authors declare no competing interests.

## References

1. Q. L. He, T. L. Hughes, N. P. Armitage, Y. Tokura, and K. L. Wang, Topological spintronics and magnetoelectronics, Nat. Mater. **21**, 15 (2022).
2. R. Tomasello, E. Martinez, R. Zivieri, L. Torres, M. Carpentieri, and G. Finoccihio, A strategy for the design of skyrmion racetrack memories, Sci. Rep. **4**, 6784 (2014).
3. X. Zhang, G. P. Zhao, H. Fangohr, J. P. Liu, W. X. Xia, and F. J. Morvan, Skyrmion-skyrmion and skyrmion-edge repulsions in skyrmion-based racetrack memory, Sci. Rep. **5**, 7643 (2015).
4. N. Nagaosa and Y. Tokura, Topological properties and dynamics of magnetic skyrmions, Nat. Nanotech. **8**, 899 (2013).
5. C. H. Lee, S. Imhof, C. Berger, F. Bayer, J. Brehm, L. W. Molenkamp, T. Kiessling, and R. Thomale, Topoelectrical Circuits, Commun. Phys. **1**, 39 (2018).
6. S. M. Rafi-Ul-Islam, Z. B. Siu, and M. B. A. Jalil, Topoelectrical circuit realization of a Weyl semimetal heterojunction, Commun. Phys. **1**, 72 (2020).
7. N. Morali, R. Batabyal, P. K. Nag, E. Liu, Q. Xu, Y. Sun, B. Yan, C. Felser, N. Avraham, and H. Beidenkopf, Fermi-arc diversity on surface terminations of the magnetic Weyl semimetal Co$_3$Sn$_2$S$_2$, Science **365**, 1286 (2019).
8. D. F. Liu *et al.*, Magnetic Weyl semimetal phase in a Kagomé crystal, Science **365**, 1282 (2019).
9. S. Y. Xu *et al.*, Discovery of a Weyl fermion semimetal and topological Fermi arcs, Science **349**, 613 (2015).

10. S. M. Huang *et al.*, A Weyl Fermion semimetal with surface Fermi arcs in the transition metal monopnictide TaAs class, Nat. Commun. **6**, 7373 (2015).
11. B. Q. Lv *et al.*, Experimental discovery of Weyl semimetal TaAs, Phys. Rev. X **5**, 031013 (2015).
12. D. F. Liu *et al.*, Magnetic Weyl semimetal phase in a Kagomé crystal, Science **365**, 1282 (2019).
13. I. Belopolski *et al.*, Discovery of topological Weyl fermion lines and drumhead surface states in a room temperature magnet, Science **365**, 1278 (2019).
14. X. Huang *et al.*, Observation of the chiral-anomaly-induced negative magnetoresistance in 3D Weyl semimetal TaAs, Phys. Rev. X **5**, 031023 (2015).
15. A. Neubauer, C. Pfleiderer, B. Binz, A. Rosch, R. Ritz, P. G. Niklowitz, and P. Böni, Topological Hall Effect in the *A* Phase of MnSi, Phys. Rev. Lett. **102**, 186602 (2009).
16. Y. Chen, D. L. Bergman, and A. A. Burkov, Weyl fermions and the anomalous Hall effect in metallic ferromagnets, Phys. Rev. B **88**, 125110 (2013).
17. K. Takiguchi *et al.*, Quantum transport evidence of Weyl fermions in an epitaxial ferromagnetic oxide, Nat. Commun. **11**, 4969 (2020).
18. J. Matsuno, N. Ogawa, K. Yasuda, F. Kagawa, W. Koshibae, N. Nagaosa, Y. Tokura, and M. Kawasaki, Interface-driven topological Hall effect in $SrRuO_3$-$SrIrO_3$ bilayer, Sci. Adv. **2**, e1600304 (2016).
19. Y. Ohuchi, J. Matsuno, N. Ogawa, Y. Kozuka, M. Uchida, Y. Tokura, and M. Kawasaki, Electric-field control of anomalous and topological Hall effects in oxide bilayer thin films, Nat. Commun. **9**, 213 (2018).
20. Q. Qin, L. Liu, W. Lin, X. Shu, Q. Xie, Z. Lim, C. Li, S. He, G. M. Chow, and J. Chen, Emergence of Topological Hall Effect in a $SrRuO_3$ Single Layer, Adv. Mater. **31**, 1807008 (2019).
21. S. Kaneta-Takada *et al.*, High-mobility two-dimensional carriers from surface Fermi arcs in magnetic Weyl semimetal films, npj Quantum Mater. **7**, 102 (2022).
22. U. Kar *et al.*, The thickness dependence of quantum oscillations in ferromagnetic Weyl metal $SrRuO_3$, npj Quantum Mater. **8**, 8 (2023).
23. S. Kaneta-Takada, Y. K. Wakabayashi, Y. Krockenberger, S. Ohya, M. Tanaka, Y. Taniyasu, H. Yamamoto, Thickness-dependent quantum transport of Weyl fermions in ultra-high-quality $SrRuO_3$ films, Appl. Phys. Lett. **118**, 092408 (2021).
24. W. Lin *et al.*, Electric Field Control of the Magnetic Weyl Fermion in an Epitaxial $SrRuO_3$ (111) Thin Film, Adv. Mater. **33**, 2101316 (2021).
25. G. Kimbell, C. Kim, W. Wu, M. Cuoco, and J. W. A. Robinson, Challenges in identifying chiral spin textures via the topological Hall effect, Commun. Mater. **3**, 19 (2022).
26. M. Cuoco and A. D. Bernardo, Materials challenges for $SrRuO_3$: From conventional to quantum electronics, APL Mater. **10**, 090902 (2022).
27. Y. Gu, Q. Wang, W. Hu, W. Liu, Z. Zhang, F. Pan, and C. Song, An overview of $SrRuO_3$-based heterostructures for spintronic and topological phenomena, J. Phys. D: Appl. Phys. **55**, 233001 (2022).
28. D. Kan, T. Moriyama, K. Kobayashi, and Y. Shimakawa, Alternative to the topological interpretation of the transverse resistivity anomalies in $SrRuO_3$, Phys. Rev. B **98**, 180408(R) (2018).
29. D. Kan and Y. Shimakawa, Defect-Induced Anomalous Transverse Resistivity in an Itinerant Ferromagnetic Oxide, Phys. Status Solidi B **255**, 1800175 (2018).

30. G. Kimbell, P. M. Sass, B. Woltjes, E. K. Ko, T. W. Noh, W. Wu, and J. W. A. Robinson, Two-channel anomalous Hall effect in SrRuO$_3$, Phys. Rev. Mater. **4**, 054414 (2020).
31. G. Kim *et al.*, Inhomogeneous ferromagnetism mimics signatures of the topological Hall effect in SrRuO$_3$ films, Phys. Rev. Mater. **4**, 104410 (2020).
32. D. Kan, T. Moriyama, and Y. Shimakawa, Field-sweep-rate and time dependence of transverse resistivity anomalies in ultrathin SrRuO$_3$ films, Phys. Rev. B **101**, 014448 (2020).
33. B. Sohn, B. Kim, J. W. Choi, S. H. Chang, J. H. Han, and C. Kim, Hump-like structure in Hall signal from ultra-thin SrRuO3 films without inhomogeneous anomalous Hall effect, Curr. Appl. Phys. **20**, 186 (2020).
34. L. Wysocki, L. Yang, F. Gunkel, R. Dittmann, P. H. M. van Loosdrecht, and I. Lindfors-Vrejoiu, Validity of magnetotransport detection of skyrmions in epitaxial SrRuO$_3$ heterostructures, Phys. Rev. Mater. **4**, 054402 (2020).
35. L. Yang, L. Wysocki, J. Schöpf, L. Jin, A. Kovács, F. Gunkel, R. Dittmann, P. H. M. van Loosdrecht, and I. Lindfors-Vrejoiu, Origin of the hump anomalies in the Hall resistance loops of ultrathin SrRuO$_3$/SrIrO$_3$ multilayers, Phys. Rev. Mater. **5**, 014403 (2021).
36. M. Zhu, Z. Cui, W. Li, Z. Shan, J. Wang, H Huang, Z. Fu, and Y. Lu, Anomalous Hall effect in spatially inhomogeneous SrRuO$_3$ films, AIP Adv. **11**, 125027 (2021).
37. Y. K. Wakabayashi, T. Otsuka, Y. Krockenberger, H. Sawada, Y. Taniyasu, and H. Yamamoto, Machine-learning-assisted thin-film growth: Bayesian optimization in molecular beam epitaxy of SrRuO$_3$ thin films, APL Mater. **7**, 101114 (2019).
38. Y. K. Wakabayashi, T. Otsuka, Y. Krockenberger, H. Sawada, Y. Taniyasu, and H. Yamamoto, Bayesian optimization with experimental failure for high-throughput materials growth, npj Comput. Mater. **8**, 180 (2022).
39. M. Naito and H. Sato, Stoichiometry control of atomic beam fluxes by precipitated impurity phase detection in growth of (Pr,Ce)$_2$CuO$_4$ and (La,Sr)$_2$CuO$_4$ films, Appl. Phys. Lett. **67**, 2557 (1995).
40. H. Yamamoto, Y. Krockenberger, Y. Taniayasu, and Y. K. Wakabayashi, Electron-Beam-Evaporation-Based Multi-Source Oxide MBE as a Synthesis Method for High-Quality and Novel Magnetic Materials−Beyond 3d Transition Metal Compounds, J. Cryst. Growth **593**, 126778 (2022).
41. H. Yamamoto, Y. Krockenberger, and M. Naito, Multi-source MBE with high-precision rate control system as a synthesis method sui generis for multi-cation metal oxides, J. Cryst. Growth **378**, 184 (2013).
42. Y. K. Wakabayashi, T. Otsuka, Y. Taniyasu, H. Yamamoto, and H. Sawada, Improved adaptive sampling method utilizing Gaussian process regression for prediction of spectral peak structures, Appl. Phys. Express **11**, 112401 (2018).
43. Y. K. Wakabayashi, Y. Krockenberger, N. Tsujimoto, T. Boykin, S. Tsuneyuki, Y. Taniyasu, and H. Yamamoto, Ferromagnetism above1000 K in a highly cation-ordered double-perovskite insulator Sr$_3$OsO$_6$, Nat. Commun. **10**, 535 (2019).
44. Y. K. Wakabayashi, S. Kaneta-Takada, Y. Krockenberger, S. Ohya, M. Tanaka, Y. Taniyasu, and H. Yamamoto, Structural and transport properties of highly Ru-deficient SrRu$_{0.7}$O$_3$ thin films prepared by molecular beam epitaxy: Comparison with stoichiometric SrRuO$_3$, AIP Advances **11**, 035226 (2021).
45. Y. K. Wakabayashi, Y. Krockenberger, T. Otsuka, H. Sawada, Y. Taniyasu, and H.

Yamamoto, Intrinsic physics in magnetic Weyl semimetal SrRuO$_3$ films addressed by machine-learning-assisted molecular beam epitaxy, Jpn. J. Appl. Phys. **62**, SA0801 (2022).
46. See Supplemental Material at xxxx.
47. Y. K. Wakabayashi, S. Kaneta-Takada, Y. Krockenberger, Y. Taniyasu, and H. Yamamoto, Wide-range epitaxial strain control of electrical and magnetic properties in high-quality SrRuO$_3$ films, ACS Appl. Electron. Mater. **3**, 2712 (2021).
48. Y. K. Wakabayashi *et al.*, Single-domain perpendicular magnetization induced by the coherent O 2*p*-Ru 4*d* hybridized state in an ultra-high-quality SrRuO$_3$ film, Phys. Rev. Mater. **5**, 124403 (2021).
49. Y. K. Wakabayashi, M. Kobayashi, Y. Takeda, M. Kitamura, T. Takeda, R. Okano, Y. Krockenberger, Y. Taniyasu, and H. Yamamoto, Isotropic orbital magnetic moments in magnetically anisotropic SrRuO$_3$ films, Phys. Rev. Mater. **6**, 094402 (2022).
50. R. A. Smith, Semiconductors, Cambridge University Press, Cambridgem UK (1978).
51. C. M. Hurd, The Hall effect in metals and alloys, Plenum Press, New York (1972).
52. G. Santi and T. Jarlborg, Calculation of the electronic structure and the magnetic properties of SrRuO$_3$ and CaRuO$_3$, J. Phys. Cond. Mat. **9**, 9563 (1997).
53. N. Haham, Y. Shperber, M. Schultz, N. Naftalis, E. Shimshoni, J. W. Reiner, and L. Klein, Scaling of the anomalous Hall effect in SrRuO$_3$, Phys. Rev. B **84**, 174439 (2011).
54. H. Kontani, T. Tanaka, and K. Yamada, Intrinsic anomalous Hall effect in ferromagnetic metals studied by the multi-*d*-orbital tight binding model, Phys. Rev. B **75**, 184416 (2007).
55. R. Karplus and J. M. Luttinger, Hall Effect in Ferromagnetics, Phys. Rev. **95**, 1154 (1954).
56. L. Berger, Side-Jump Mechanism for the Hall Effect of Ferromagnets, Phys. Rev. B **2**, 4559 (1970).
57. L. Berger, Application of the Side-Jump Model to the Hall Effect and Nernst Effect in Ferromagnets, Phys. Rev. B **5**, 1862 (1972).
58. J. Smit, The spontaneous hall effect in ferromagnetics I, Physica **21**, 877 (1955).
59. J. Smit, The spontaneous hall effect in ferromagnetics II, Physica **24**, 39 (1958).
60. N. Haham, J. W. Reiner, and L. Klein, Scaling of the paramagnetic anomalous Hall effect in SrRuO$_3$, Phys. Rev. B **86**, 144414 (2012).

**Figures and figure captions**

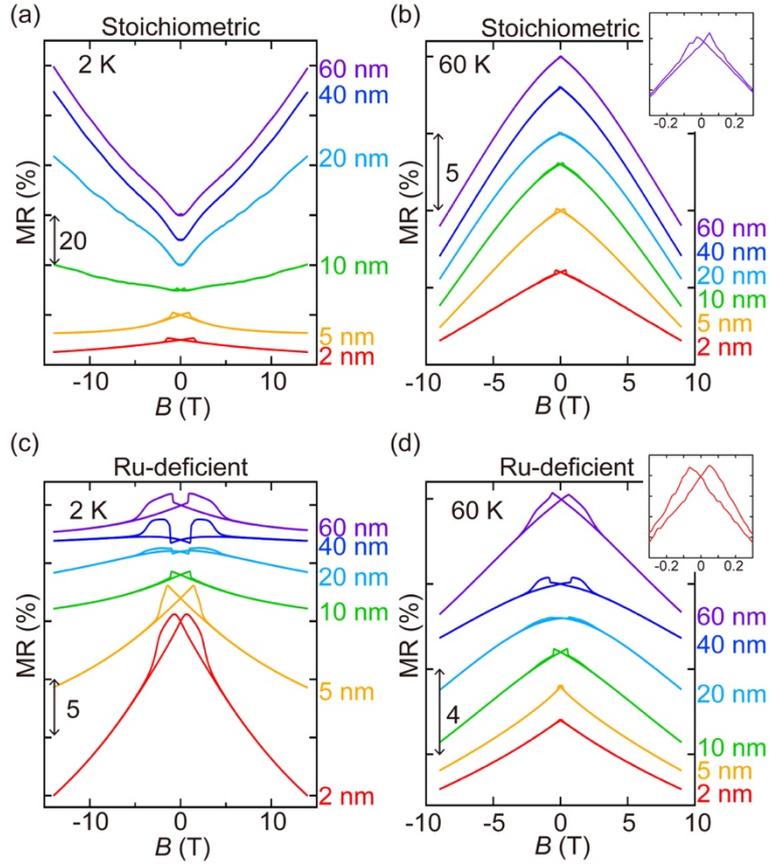

FIG. 1. (a)−(d) Thickness $t$ dependence of the MR for the stoichiometric $SrRuO_3$ [(a), (b)] and the Ru-deficient $SrRu_{0.7}O_3$ [(c), (d)] films at 2 K [(a),(c)] and 60 K [(b),(d)], respectively. $B$ is applied in the out-of-plane [001] direction of the $SrTiO_3$ substrate. The insets of Figs. 1(b) and (d) are the enlarged AMR of the $SrRuO_3$ film with $t$ = 60 nm and the $SrRu_{0.7}O_3$ film with $t$ = 2 nm, respectively.

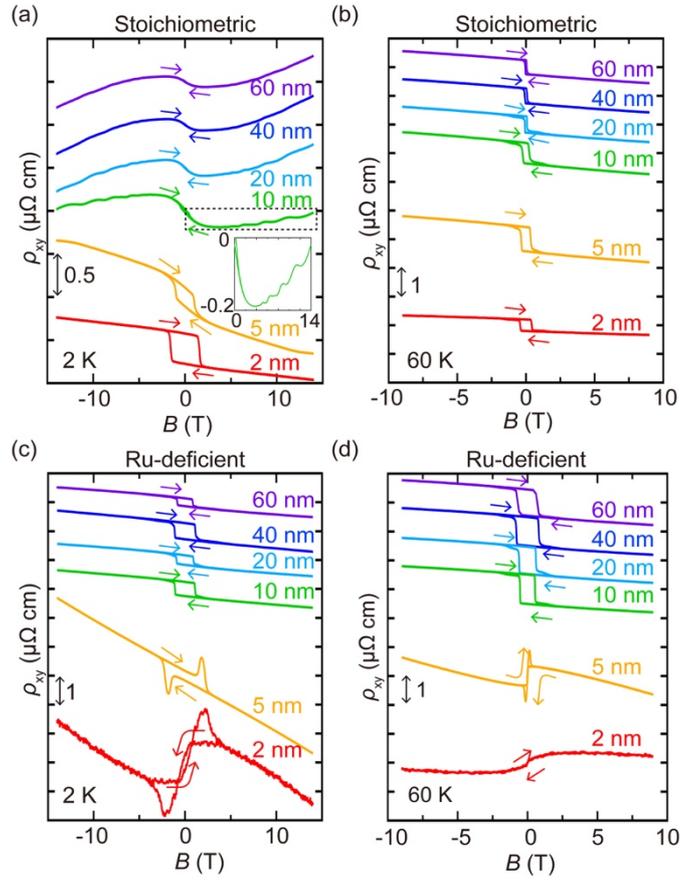

FIG. 2. (a)−(d) Thickness $t$ dependence of the Hall resistivity $\rho_{xy}$ for the stoichiometric SrRuO$_3$ [(a), (b)] and the Ru-deficient SrRu$_{0.7}$O$_3$ [(c), (d)] films at 2 K [(a),(c)] and 60 K [(b),(d)], respectively. $B$ is applied in the out-of-plane [001] direction of the SrTiO$_3$ substrate. The arrows indicate the sweep direction of $B$.

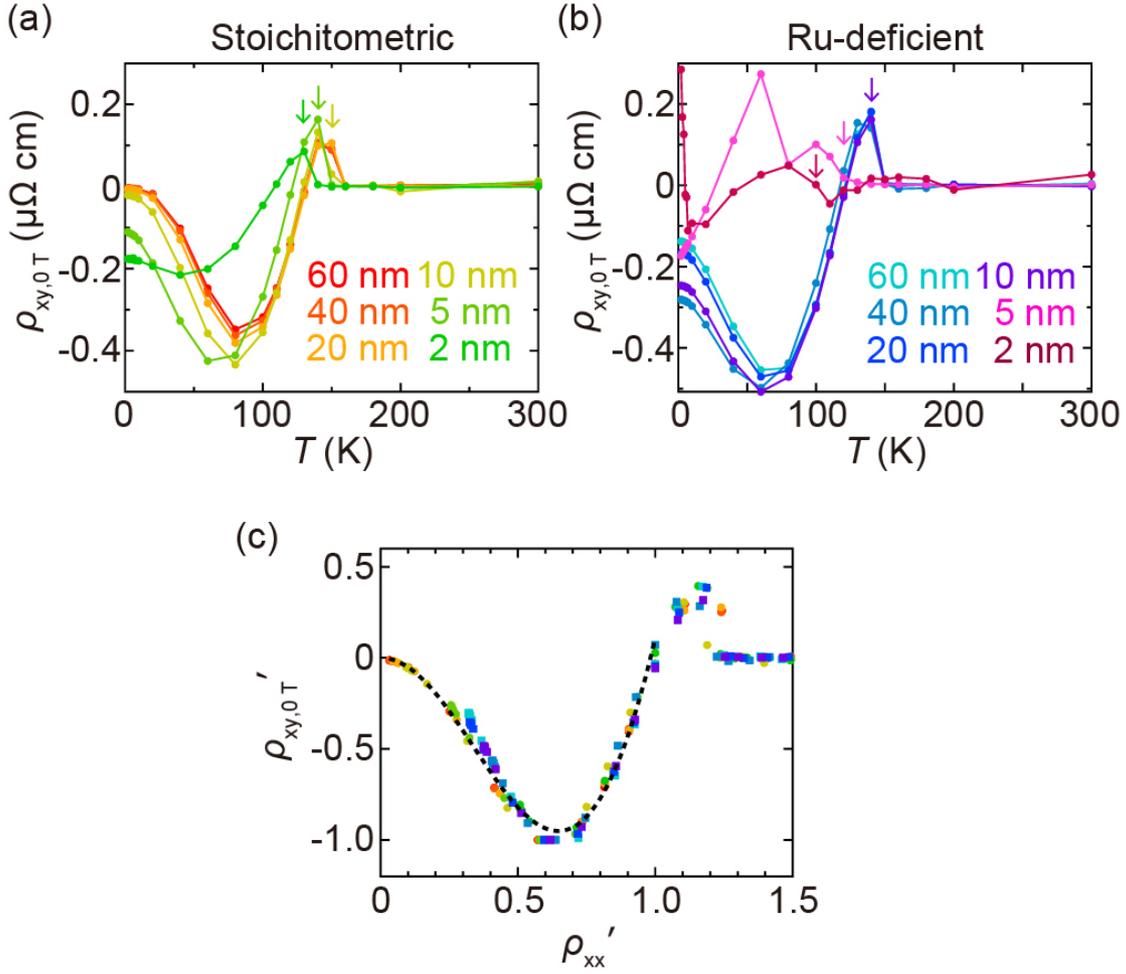

FIG. 3. (a), (b) Remanent Hall resistivity $\rho_{xy,0\,T}$ vs. temperature $T$ curves for the stoichiometric SrRuO$_3$ (a) and Ru-deficient SrRu$_{0.7}$O$_3$ (b) films with $t = 2-60$ nm. (c) Normalized remanent Hall resistivity $\rho_{xy,\,0\,T}'$ vs. normalized longitudinal resistivity $\rho_{xx}'$ curves for the stoichiometric SrRuO$_3$ and Ru-deficient SrRu$_{0.7}$O$_3$ films, excluding the Ru-deficient SrRu$_{0.7}$O$_3$ films with $t = 2$ and $5$ nm. The dashed line is the fitting of a scaling law by Eq. (1).

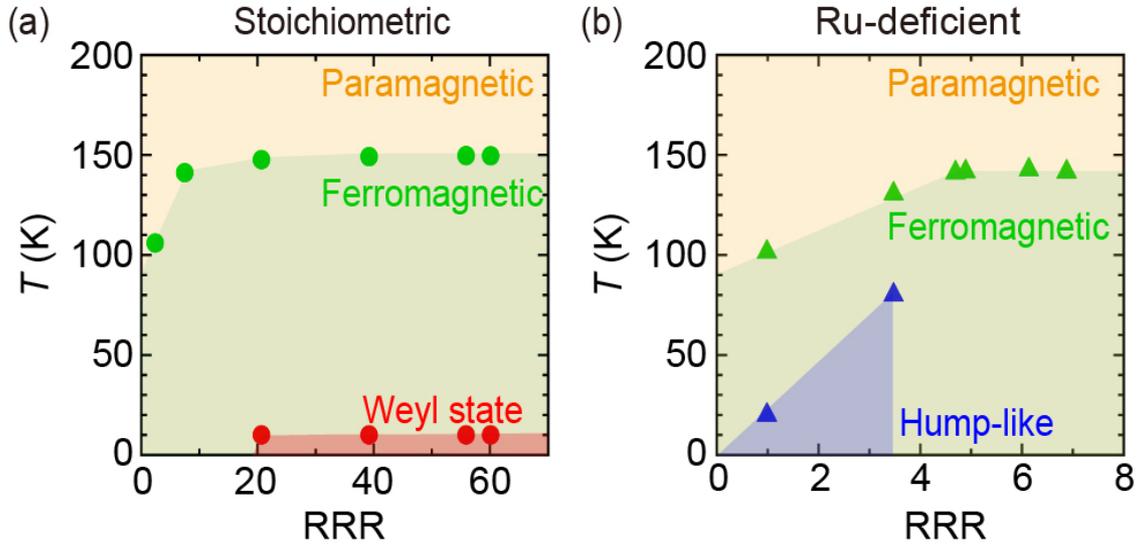

FIG. 4. RRR and temperature dependence of the transport phenomena in the stoichiometric SrRuO$_3$ (a) and Ru-deficient SrRu$_{0.7}$O$_3$ (b) films. Here, red, blue, and green plots are the highest temperature where the Weyl fermion transport appears, the highest temperature where the hump-like Hall effect anomaly appears, and $T_C$. The pale-red and pale-blue areas roughly correspond to the clean and dirty regimes, respectively, defined based on the results of the Hall effect experiments (see text).

# Supplemental Material for

# Scattering-dependent transport of SrRuO$_3$ films: From Weyl fermion transport to hump-like Hall effect anomaly

Shingo Kaneta-Takada,[1,2,*] Yuki K. Wakabayashi,[1,*,†] Yoshiharu Krockenberger,[1] Hiroshi Irie,[1] Shinobu Ohya,[2,3] Masaaki Tanaka,[2,3] Yoshitaka Taniyasu,[1] and Hideki Yamamoto[1]

[1]*NTT Basic Research Laboratories, NTT Corporation, Atsugi, Kanagawa 243-0198, Japan*
[2]*Department of Electrical Engineering and Information Systems, The University of Tokyo, Bunkyo, Tokyo 113-8656, Japan*
[3]*Center for Spintronics Research Network (CSRN), The University of Tokyo, Bunkyo, Tokyo 113-8656, Japan*

[*]These authors contributed equally to this work.
[†]Corresponding author: yuuki.wakabayashi.we@hco.ntt.co.jp

# I. Transport characteristics of stoichiometric SrRuO$_3$ and Ru-deficient SrRu$_{0.7}$O$_3$ films

Figure S1 shows the longitudinal resistivity $\rho_{xx}(T)$ of the specimens processed into the Hall bar structure. The $\rho_{xx}(T)$ curves show clear kinks around arrows [Figs. S1(a) and S1(b)], at which the ferromagnetic transition occurs and spin-dependent scattering is suppressed. To highlight the ferromagnetic transition, we plot the derivative resistivity $d\rho_{xx}/dT$ as a function of $T$ [Figs. S1(c) and S1(d)]. Here, we define $T_C$ as the temperature at which the $d\rho_{xx}/dT$ curves show peaks or kinks. The thickness $t$ dependence of $T_C$ is shown in Fig. S1(f). We also estimate the residual resistivity ratio (RRR) [=$\rho_{xx}(300\ K)/\rho_{xx}(2\ K)$] from Fig. S1(a) and S1(b) [Fig. S1(e)].

Compared with the specimens before the Hall-bar fabrication, temperature dependencies of $\rho_{xx}(T, 0\ T)$ were almost preserved for the SrRuO$_3$ films with $t = 5 - 60$ nm and SrRu$_{0.7}$O$_3$ films, while the absolute values became ~ 1.3 times higher likely due to reduction of effective cross-sectional area of the Hall bar channel through the fabrication process. This confirms the uniformity of the original specimens and limited fabrication damage, in addition to the high reproducibility of our experiments. In contrast, in the extremely thin films, the influence of the Hall-bar fabrication is not negligible: the $\rho_{xx}(T, 0\ T)$ values of the 2-nm-thick SrRuO$_3$ film became substantially smaller, which might be explained by the formation of a conducting portion in the SrTiO$_3$ substrate due to Ar ion milling. Besides, the $\rho_{xx}(T, 0\ T)$ values of the 2-nm and 5-nm thick SrRu$_{0.7}$O$_3$ films respectively became 1.6 – 2 and ~7 times larger, which might be explained by the reduction of the Hall bar channel cross-sectional area from its designed value during the fabrication process. Since there is no significant change in the RRR and $T_C$ values before and after the Hall-bar fabrication, the fabrication process had little effect on the electronic state of SrRuO$_3$ and SrRu$_{0.7}$O$_3$.

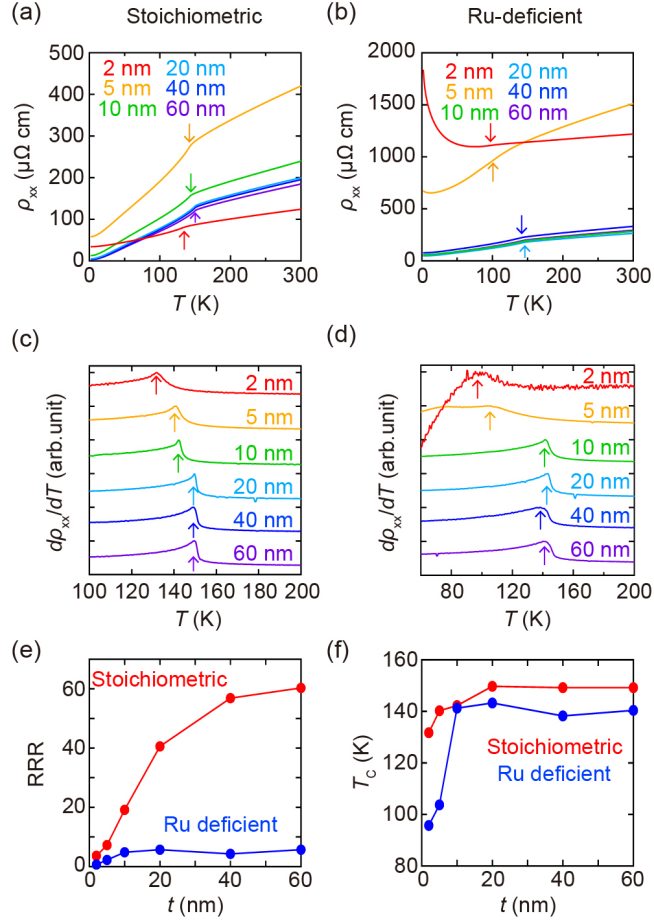

FIG. S1. (a), (b) Temperature dependence of longitudinal resistivity for the (a) stoichiometric SrRuO$_3$ and (b) Ru-deficient SrRu$_{0.7}$O$_3$ films with various thicknesses $t$ of 2–60 nm. Arrows indicate the kink positions, which are the same positions as those in (c) and (d). (c), (d) Temperature dependence of differential resistivity for the (c) stoichiometric SrRuO$_3$ and (d) Ru-deficient SrRu$_{0.7}$O$_3$ films with various thicknesses $t$ of 2–60 nm. Arrows indicate the peak or kink positions in the $d\rho_{xx}/dT$ curves. (e) Thickness dependence of the RRR for the stoichiometric SrRuO$_3$ and Ru-deficient SrRu$_{0.7}$O$_3$ films. (f) Thickness dependence of $T_C$ obtained from the peak or kink positions in differential resistivity in (c) and (d).

## II. Magnetotransport of the stoichiometric SrRuO$_3$ and Ru-deficient SrRu$_{0.7}$O$_3$ films above $T_C$

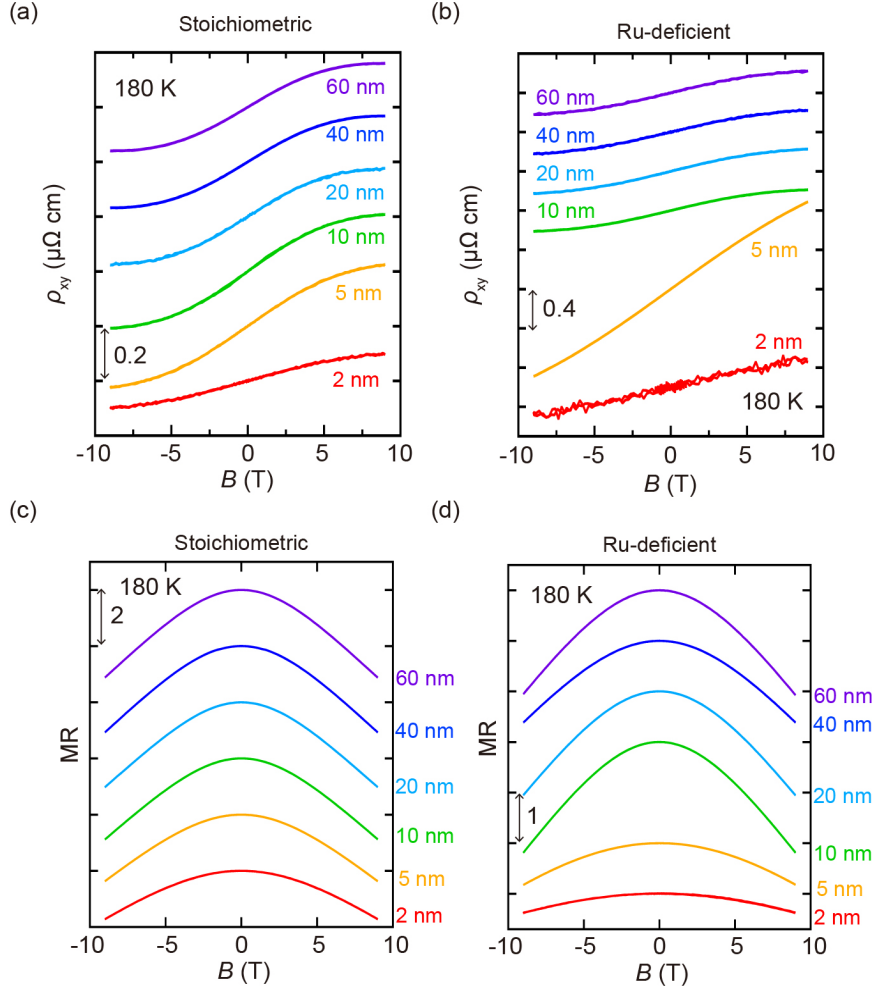

FIG. S2. (a), (b) Thickness $t$ dependence of the Hall resistivity $\rho_{xy}$ for (a) the stoichiometric SrRuO$_3$ and (b) the Ru-deficient SrRu$_{0.7}$O$_3$ films measured at 180 K ($> T_C$), respectively. (c), (d) Thickness $t$ dependence of the MR for (c) the stoichiometric SrRuO$_3$ and (d) the Ru-deficient SrRu$_{0.7}$O$_3$ films measured at 180 K ($> T_C$), respectively. Here, the temperature of 180 K ($> T_C$) was chosen to observe $\rho_{xy}$ and MR in the paramagnetic states. $B$ was applied in the out-of-plane [001] direction of the SrTiO$_3$ substrate.

## III. Thickness $t$ dependence of the carrier density and mobility for SrRuO$_3$ and Ru-deficient SrRu$_{0.7}$O$_3$ films under an assumption of the Drude model with one majority carrier

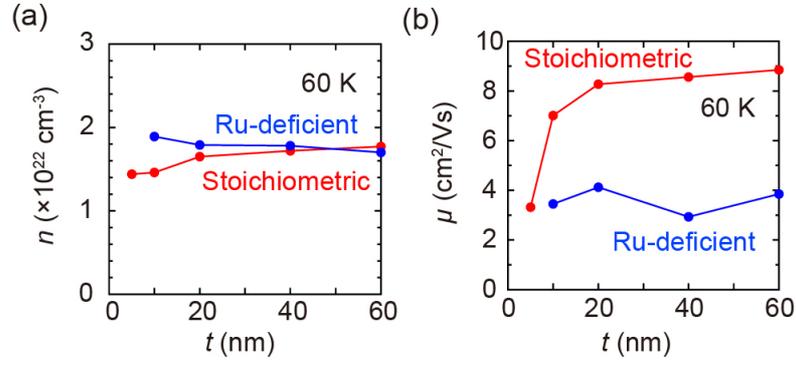

FIG. S3. Thickness $t$ dependence of the carrier density (a) and mobility (b) at 60 K for the stoichiometric SrRuO$_3$ and Ru-deficient SrRu$_{0.7}$O$_3$ films, estimated under an assumption of the Drude model with one majority carrier (see main text).

## IV. The definition of $\rho_{xy,0\,T}$

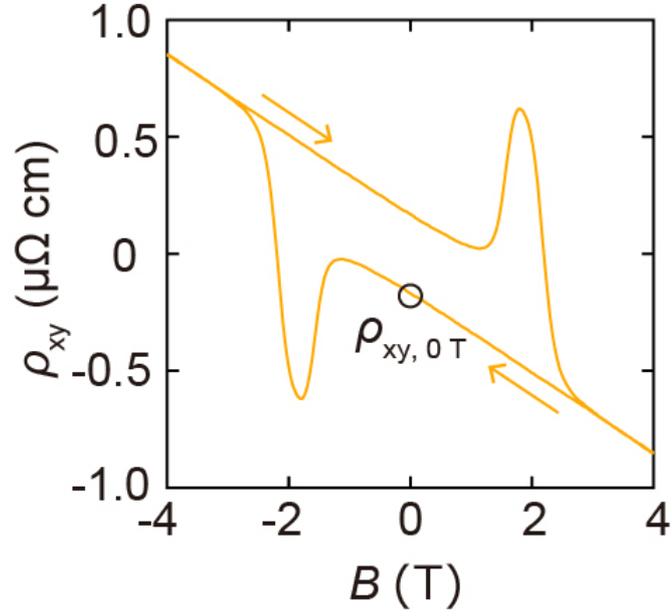

FIG. S4. Magnetic field $B$ dependence of Hall resistivity $\rho_{xy}$ for the Ru-deficient SrRu$_{0.7}$O$_3$ films with $t = 5$ nm at 2 K. The circle at 0 T is $\rho_{xy,\,0\,T}$ after sweeping from high $B$ to 0 T.

**V. The $\rho_{xy, 0\,T}'$ vs. $\rho_{xx}'$ curves for all the SrRuO$_3$ films and Ru-deficient SrRu$_{0.7}$O$_3$ films**

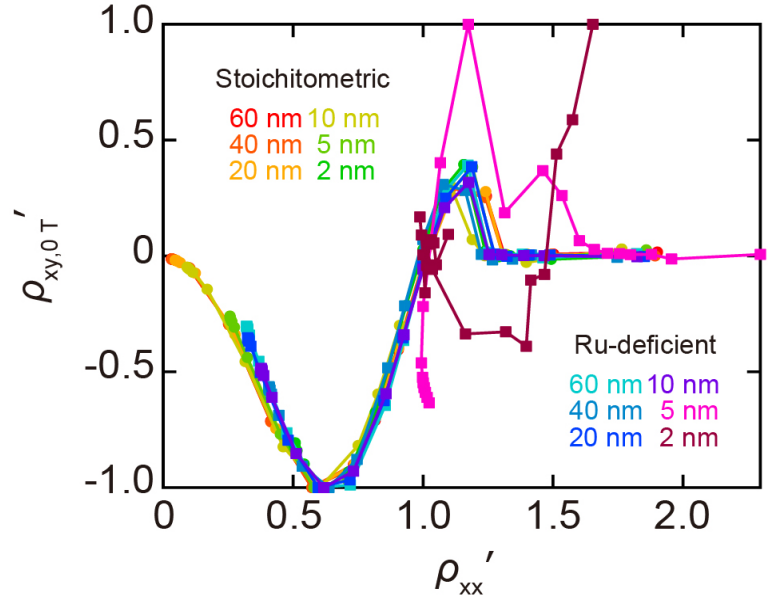

FIG. S5. Normalized remanent Hall resistivity $\rho_{xy,\,0\,T}'$ vs. normalized longitudinal resistivity $\rho_{xx}'$ curves for the stoichiometric SrRuO$_3$ and Ru-deficient SrRu$_{0.7}$O$_3$ films with $t = 2-60$ nm.